\documentclass[prl,twocolumn,showpacs,preprintnumbers,amsmath,amssymb,superscriptaddress]{revtex4}
\usepackage{graphicx}
\begin{document}

\title{On the nature of the phase transition in the itinerant helimagnet MnSi}

\author{S. M. Stishov}
\email{sergei@hppi.troitsk.ru}\affiliation{Institute for High
Pressure Physics, Troitsk, Moscow Region, Russia}
\author{A.E. Petrova}
\affiliation{Institute for High Pressure Physics, Troitsk, Moscow
Region, Russia}
\author{S. Khasanov}
\affiliation{Institute of Solid State Physics, Chernogolovka,
Moscow Region, Russia}
\author{G. Kh. Panova}
\affiliation{Russian Research Center Kurchatov Institute, Moscow,
Russia}
\author{A.A.Shikov}
\affiliation{Russian Research Center Kurchatov Institute, Moscow,
Russia}
\author{J. C. Lashley}
\affiliation{Los Alamos National Laboratory, Los Alamos, 87545 NM,
USA}
\author{D. Wu}
\affiliation{Ames Laboratory, Iowa State University, Ames, IA
50011, USA}
\author{T. A. Lograsso}
\affiliation{Ames Laboratory, Iowa State University, Ames, IA
50011, USA}

\date{\today}

\begin{abstract}

A careful study of thermodynamic and transport properties of a
high quality single crystal of MnSi at ambient pressure suggests
that its transition to a helical magnetic state near 29 K is
weakly first order. The heat capacity, temperature derivative of
resistivity, thermal expansion and magnetic susceptibility exhibit
a specific structure around the phase transition point,
interpreted as a combination of first and second order features.
Conclusions drawn from these experiments question prevailing views
on the phase diagram of MnSi and propose that the phase transition
under study becomes second order at high pressure and low
temperature.
\end{abstract}


\maketitle

Physical properties of the cubic intermetallic compound MnSi, a
weak itinerant helimagnet with Dzyaloshinski - Moria interaction
\cite{1,2} have been studied extensively for more than for 40
years since discovery of its spin ordered phase at the temperature
slightly below 30 K \cite{3}. Neutron-diffraction measurements
showed that MnSi had a helical spin structure with a propagation
vector in the [111] direction \cite{4}. Interest in studying of
MnSi was greatly enhanced by the finding that the temperature of
the magnetic phase transition decreased with pressure and tended
to zero at about 1.4 GPa with expectations of the quantum critical
behavior \cite{5}.

Despite theoretical conclusions that phase transition in MnSi can
be fluctuation-induced first-order \cite{6,7,8} all the physical
properties of MnSi studied up to now seem to be continuous across
the phase-transition line at ambient pressure. Consequently, when
a significant change was observed in the temperature dependence of
ac susceptibility at the phase transition under high pressure was
found \cite{9,10} (see, also \cite{11,12}), it was accepted as a
manifestation of the existence of a tricritical point and
first-order nature of the phase transition in MnSi at low
temperatures and high pressures. Nevertheless, some remarkable
properties of the phase transition in MnSi are not understood. In
particular, some quantities, like thermal expansion coefficient
\cite{13}, heat capacity \cite{14}, temperature coefficient of
resistivity \cite{12} display a well-defined shoulders on the high
temperature side of their corresponding peaks at the phase
transition and the nature of these shoulders remains a puzzle
\cite{15}.With these unresolved issues, it is appropriate to carry
out systematic investigation of physical properties of MnSi with
the same well-characterized sample.

\begin{figure}[htb]
\includegraphics[width=86mm]{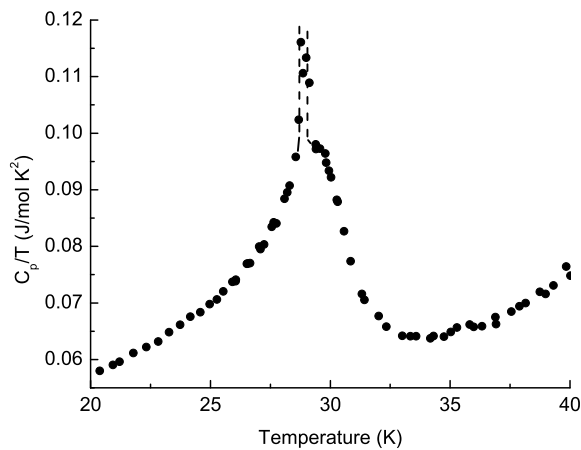}
\caption{\label{fig1} Temperature dependence of heat capacity
divided by temperature near the phase transition in MnSi  }
\end{figure}

To this end, heat capacity, electrical resistivity, thermal
expansion in magnetic fields, dc and ac magnetic susceptibility
were measured at ambient pressure on a high quality MnSi single
crystal. As will be discussed, the heat capacity and temperature
derivative of resistivity, thermal expansion and magnetic
susceptibility behave as if they diverge at the Curie point. Some
of these quantities (heat capacity, thermal expansion coefficient,
and temperature coefficient of resistivity) display a doublet
structure, with sharp and broad components at the phase transition
point. We argue that the sharp component is a slightly broadened
delta function, corresponding to a first-order phase transition.
These new results suggest that the magnetic phase transition in
MnSi is weakly first order at ambient and low pressure, with small
volume change  $\Delta V/V$ of order $10^{-6}$, and becomes second
order at pressures higher than 0.35 GPa, an interpretation that
reverses the prevailing view about the nature of the phase
transition in MnSi and that suggests a new look at the physics of
MnSi at temperatures close to zero and high pressure.
Specifically, these conclusions suggest that the asymptotic
crossing point of the phase transition line and the pressure axis
at T=0 in MnSi is a real, continuous quantum-critical point.
\begin{figure}[htb]
\includegraphics[width=86mm]{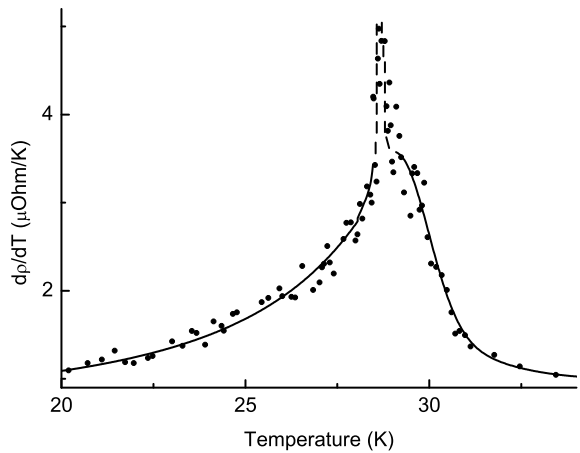}
\caption{\label{fig2} Dependence of temperature coefficient of
resistivity $d\rho/dT$ on temperature in the vicinity the phase
transition in MnSi. }
\end{figure}

For these studies, a large single crystal of MnSi was grown by the
Bridgman technique from a stoichiometric melt of appropriate
quantities of distilled manganese \cite{17} and silicon with
purities of 99.99\% and 99.999\% respectively. Samples of
necessary size and orientation for various experiments were cut by
low power spark erosion. X-ray studies give a lattice parameter $a
= 4.5598(2)$ $\AA$  at 298 K and overall mosaicity less than
$0.1^{\circ}$. A further indication of crystal quality is
reflected in a resistivity ratio $R_{300}/R_{T\rightarrow0}$ equal
to 230. From the saturated magnetization at high field and T=5K,
magnetic moment per atom Mn is 0.4  $\mu_{B}$, whereas, fitting
low field (100 Oe) inverse susceptibility data in the range
120-300 K to a Curie-Weiss form gives an effective moment of 2.27
$\mu_{B}$ per Mn in the paramagnetic phase. These values agree
well with previous reports. According to results from the current
experiments, the temperature of the phase transition in our sample
of MnSi is confined to the limits 28.7-29 K.
\begin{figure}[htb]
\includegraphics[width=86mm]{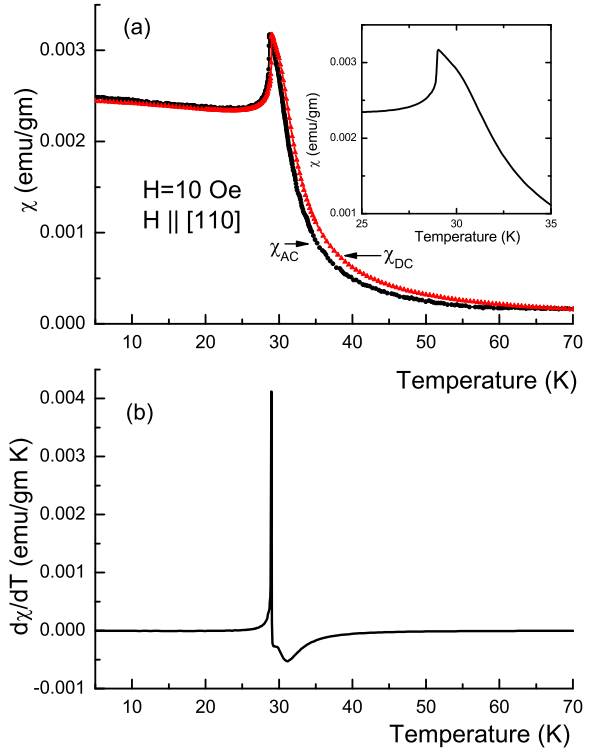}
\caption{\label{fig3} Magnetic susceptibility $\chi=M/H$ of MnSi
as function of temperature. Measurements were made in a field of
10 Oe. The ac susceptibility data were scaled by simple
multiplication. }
\end{figure}

Resistivity measurements were carried out by a standard four-probe
technique. To avoid complications connected with possible
contamination of the sample surface during cutting, polishing and
etching, we use a small splinter of the MnSi crystal with
dimensions of about $2\times0.7\times0.5\ mm^{3}$. Heat capacity
was measured by adiabatic calorimeter with an accuracy about of
1\% in the temperature interval 4-40 K. The indicated accuracy can
be achieved only with temperature steps no less than 0.5 K, and to
increase the resolution several subsequent runs were performed. DC
magnetic susceptibility measurements were made in a Quantum Design
Magnetic Properties Measurement System; whereas, ac susceptibility
was measured with a two-coil set up (drive and pick up coils) by a
standard modulation technique at a modulation frequency 19 Hz.
Linear thermal expansion measurements were performed in a
capacitance dilatometer with resolution about 0.05 $\AA$
\cite{18}. In all these experiments, temperature was measured by
the calibrated Cernox thermometers with potential overall
resolution and accuracy not worse than 0.05 K.
\begin{figure}[htb]
\includegraphics[width=86mm]{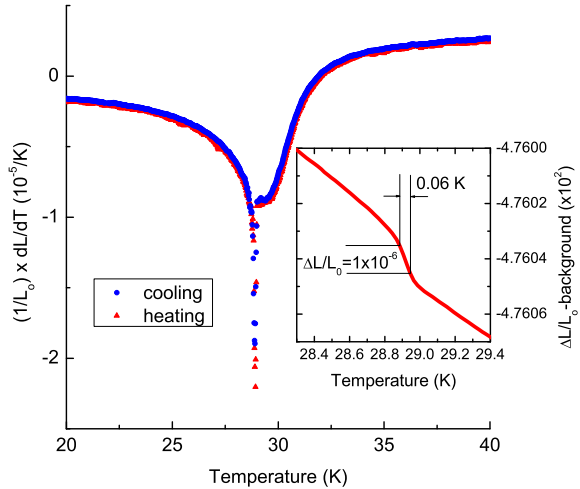}
\caption{\label{fig4} Linear thermal expansion coefficient of MnSi
in the vicinity of the phase transition. A small hysteresis about
0.03 K that can be seen at the transition in the plot
magnification should be taken into account, but probably it is too
small to ascribe to it any significant physical meaning. Variation
of the relative length of the sample close to the transition point
is shown in the inset. For better view a background contribution
was subtracted from the original data. The width of the transition
is compatible with the temperature resolution of the experiment. }
\end{figure}

\begin{figure}[htb]
\includegraphics[width=86mm]{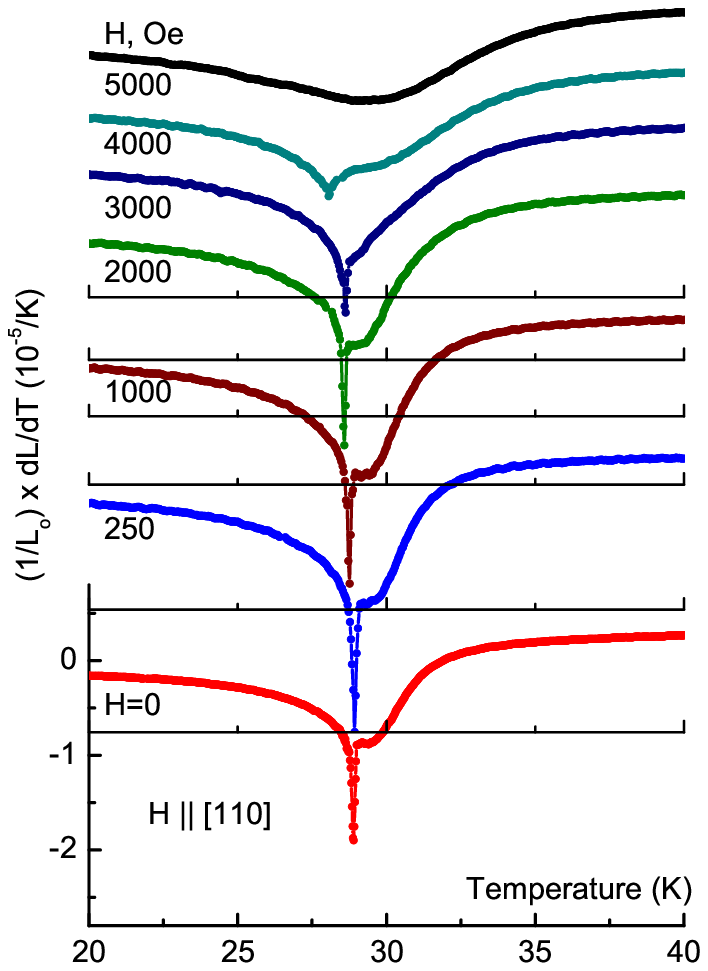}
\caption{\label{fig5} Linear thermal expansion coefficients of
MnSi near the phase transition in magnetic fields. }
\end{figure}

\begin{figure}[htb]
\includegraphics[width=86mm]{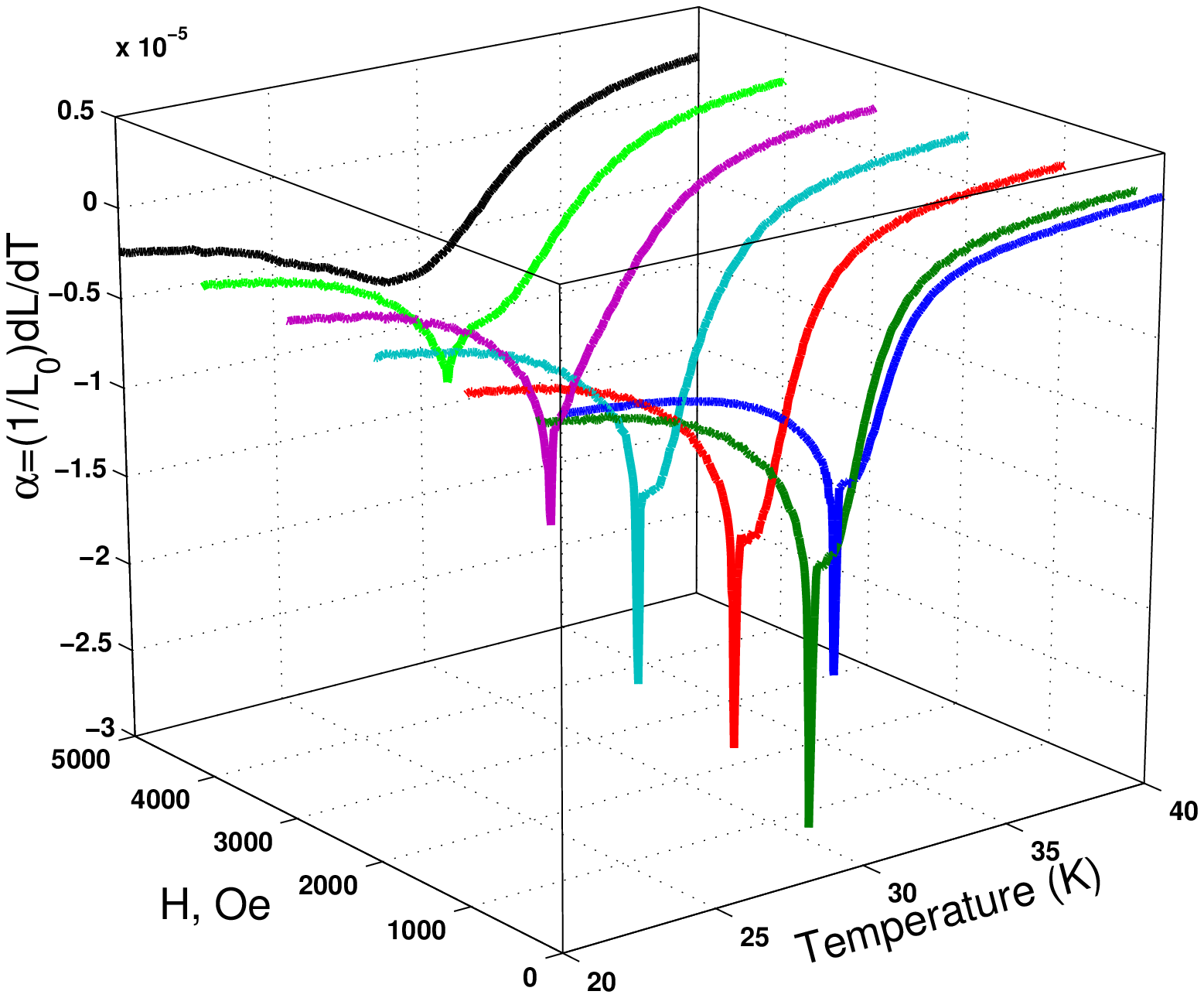}
\caption{\label{fig6} 3D plot of linear thermal expansion
coefficients of MnSi near the phase transition in magnetic fields.
}
\end{figure}

As seen in Fig.\ref{fig1} and Fig.\ref{fig2}, the heat capacity
and the temperature coefficient of resistivity of this crystal of
MnSi show well-defined shoulders on the high temperature side of
the peaks that define the phase transition. This is similar to
previous reports \cite{12,14}. A new and quite significant feature
observed in the current heat capacity data, even given the
accuracy and temperature resolution ($\sim0.2$ K) of the
experiments, is the sharp form of the prominent peak
(Fig.\ref{fig1}). The temperature coefficient of resistivity shows
the same trend within the greater scattering of the data
(Fig.\ref{fig2}).

DC and ac magnetic susceptibility data are plotted in
Fig.\ref{fig3} (a) and the temperature derivative is given in
Fig.\ref{fig3} (b). The unusual shape of  $\chi (T)$ in the
vicinity of the phase transition is seen in the inset of (a). This
form can be considered as the result of a sudden jump-like
increase in the magnetic susceptibility at the transition point,
which well demonstrated in the Fig.\ref{fig3} (b). Without this
jump, the magnetic susceptibility curve would look like one
typical of antiferromagnetic phase transition.

These indications for the first-order component of the phase
transition in MnSi are reinforced the high resolution thermal
expansion data (Fig.\ref{fig4}). The existence of slightly
broadened delta function developing within the continuous anomaly
of the thermal expansion coefficient $\alpha=(1/L_{0})dL/dT$
clearly exposes the first-order nature of the phase transition in
MnSi at ambient pressure. Integration of the curve $\alpha(T)$
(inset of Fig.\ref{fig4}) permits an estimate the relative volume
change at the first-order phase transition in MnSi as
$\sim3\times10^{-6}$.

Now we turn to Fig.\ref{fig5} and Fig.\ref{fig6}that illustrates
the influence of magnetic field on the thermal expansion
coefficient of MnSi. In case of the helical spin ordering,
magnetic field is not directly coupled with the order parameter,
and hence, no significant effect of magnetic field on the phase
transition is expected until the field-induced ferromagnetic state
appears at about 3500 Oe \cite{19}. As seen in Fig.\ref{fig6},
moderate magnetic fields up to 250 Oe even sharpen the phase
transition (probably a result of forming a single domain sample)
and obviously do not change its nature for fields less than 4000
Oe, as anticipated. We point out that any possible effects of
reorientations of the helical structure and emerging conical
magnetic structure do not reveal themselves in Fig.\ref{fig5},
Fig.\ref{fig6} but the fast degradation of the first order
transition between 3000-4000 Oe clearly indicates formation of the
ferromagnetic spin structure.

These results lead to the conclusion that the doublet structure of
the peaks of the heat capacity, temperature coefficient of
resistivity, thermal expansion coefficient and the jump in the
magnetic susceptibility at the Curie point in MnSi are result from
the combination of second-order and first-order features. This
identifies the phase transition as weakly first order, most likely
induced by fluctuations \cite{6,7,8,20}. Taking into account data
\cite{12}, where disappearance of the doublet structure in
$d\rho/dT$ at pressure was observed above 0.35 GPa, we conclude
that at higher pressures the transition becomes second order and
continues this way down to zero temperature. Consequently, the
indicated pressure 0.35 GPa should correspond to location of a
tricritical point. The above conclusions completely reverse the
widely accepted view of the phase diagram of MnSi and affirm that
the real, continuous quantum-critical point should be situated at
the transition line at $T\rightarrow0$ \cite{22,25}.
\begin{acknowledgments}
Authors are grateful to I.E. Dzyaloshinski, S.V. Maleev, S.A.
Brazovsky and J.D. Thompson for reading the manuscript and
valuable remarks. Technical assistance of J.D. Thompson, V.
Sidorov and V. V. Krasnorussky is greatly appreciated. DW and TAL
wish to acknowledge the support of the U.S. Department of Energy,
Basic Energy Sciences. SMS and AEP appreciate support of the
Russian Foundation for Basic Research (grant 06-02-16590), Program
of the Physics Department of RAS on Strongly Correlated Systems
and Program of the Presidium of RAS on Physics of Strongly
Compressed Matter. Work at Los Alamos was performed under the
auspices of the US Department of Energy, Office of Science.
\end{acknowledgments}


\begin{thebibliography}{99}
\bibitem{15} These features have been attributed to a skyrmion texture in the paramagnetic phase of MnSi
\cite{16}.
\bibitem{20} Striction effects also might be  relevant to this case \cite{21}
\bibitem{22} Observation of the inhomogeneity \cite{23,24} and phase
separation \cite{24} in MnSi at high pressure, possibly caused by
a nonhydrostatic environment, would  prevent one from detailed
study of the quantum-critical region. Development of a new
experimental approach for exploration of the quantum critical
phenomena at high pressure seems to be quite compelling.
\bibitem{25} In principal two tricritical points may exist on the transition line for materials like MnSi; however, in this case a competing spin order should be observed \cite{7}.
\end{thebibliography}

\end{document}